\documentclass[
    ,final            
  ]
  {aipproc}
\providecommand{\eprint}[2][]{\url{#2}}
\layoutstyle{6x9}

\begin{document}

\title{Search for RS-gravitons at CDF}


\author{John Strologas}{
  address={Department of Physics and Astronomy, University of New Mexico, \\Albuquerque, NM 87131, USA}
}

\begin{abstract}
We present a search for Randall-Sundrum (RS) gravitons decaying to diphotons or dielectrons or dimuons, performed with the CDF II detector and using up to 5.7 fb$^{-1}$ of integrated luminosity.  The respective mass spectra are consistent with the ones expected by the standard model.  For the RS-model parameter $k/\overline{M}_{Pl}=0.1$, RS-gravitons with mass less than 1111 GeV$/c^2$ are excluded at 95\% CL.
\end{abstract}

\maketitle


\section{Introduction}
Although extremely successful, the standard model (SM) of particles and fields is not sufficient for solving many open physics problems 
including the source of the dark matter, the incorporation of gravity, and the hierarchy problem between the weak and Planck scales.  
One theory that addresses the hierarchy problem is the Randall-Sundrum (RS) graviton model \cite{RS}.
  
The RS model solves the hierarchy problem by introducing an extra compact dimension accessible 
only to gravity.   The phenomenology leads to a small number of distinct Kaluza-Klein states (spin-2 gravitons) that 
couple with gravitational-strength to SM particles and can be detected as resonances of pairs of jets, leptons,
photons, or $W/Z$ bosons.   
Given the considerable dijet background in a hadron-collider environment, and
the low leptonic branching fraction of the weak bosons, the prompt dilepton and diphoton final states 
offer the greatest sensitivity at the Tevatron.  We present here the recent CDF searches for RS-gravitons decaying to diphotons, dielectrons, or dimuons.

\section{Search for RS-gravitons decaying to diphotons}

For the 5.4 fb$^{-1}$ diphoton search \cite{RS_gg} a combination of two diphoton ($E_T>12,18$ GeV)
and two single-photon ($E_T>50,70$ GeV) triggers is used.  
The two photon candidates are required to be central and fiducial in the detector,
have $E_T>15$ GeV and excess energy around a cone of $\Delta R = \sqrt{\Delta\phi)^2+(\Delta\eta)^2}=0.4$
less than 2 GeV.

The expected number of RS graviton events, as a function of graviton mass, is
estimated using the {\sc pythia} \cite{pythia} Monte Carlo (MC) event generator, with {\sc cteq5l} \cite{cteq} parton distribution
functions (PDF), and processed by the {\sc geant}-based \cite{geant} CDF II detector
simulation. 

The main SM diphoton background to the graviton signal is the $\gamma\gamma$ production,
determined using {\sc diphox} \cite{diphox} next-to-leading order (NLO) MC event generator, with 
acceptance corrected using a {\sc pythia} SM diphoton sample passing the CDF detector simulation, and with the
high-mass spectrum extrapolated.  A secondary background comes from jet+$\gamma$ or dijet events where one or two jets
are misidentified as photons (fake-photon background).  This background is determined
with the construction of a CDF ``fake diphoton'' sample with looser photon identification
requirements, to allow for jet contamination.  Subsequently, the actual CDF diphoton spectrum
is fitted to this ``fake diphoton'' shape along with the MC-estimated SM diphoton spectrum, with both components varying in the fit.

Systematic uncertainties arise from the luminosity determination (6\%), the initial and final 
radiation (4-8\%), and the $Q^2$ and PDF uncertainties (20\%) in the {\sc diphox} calculation.

\begin{figure}[t]
\includegraphics[scale=.4]{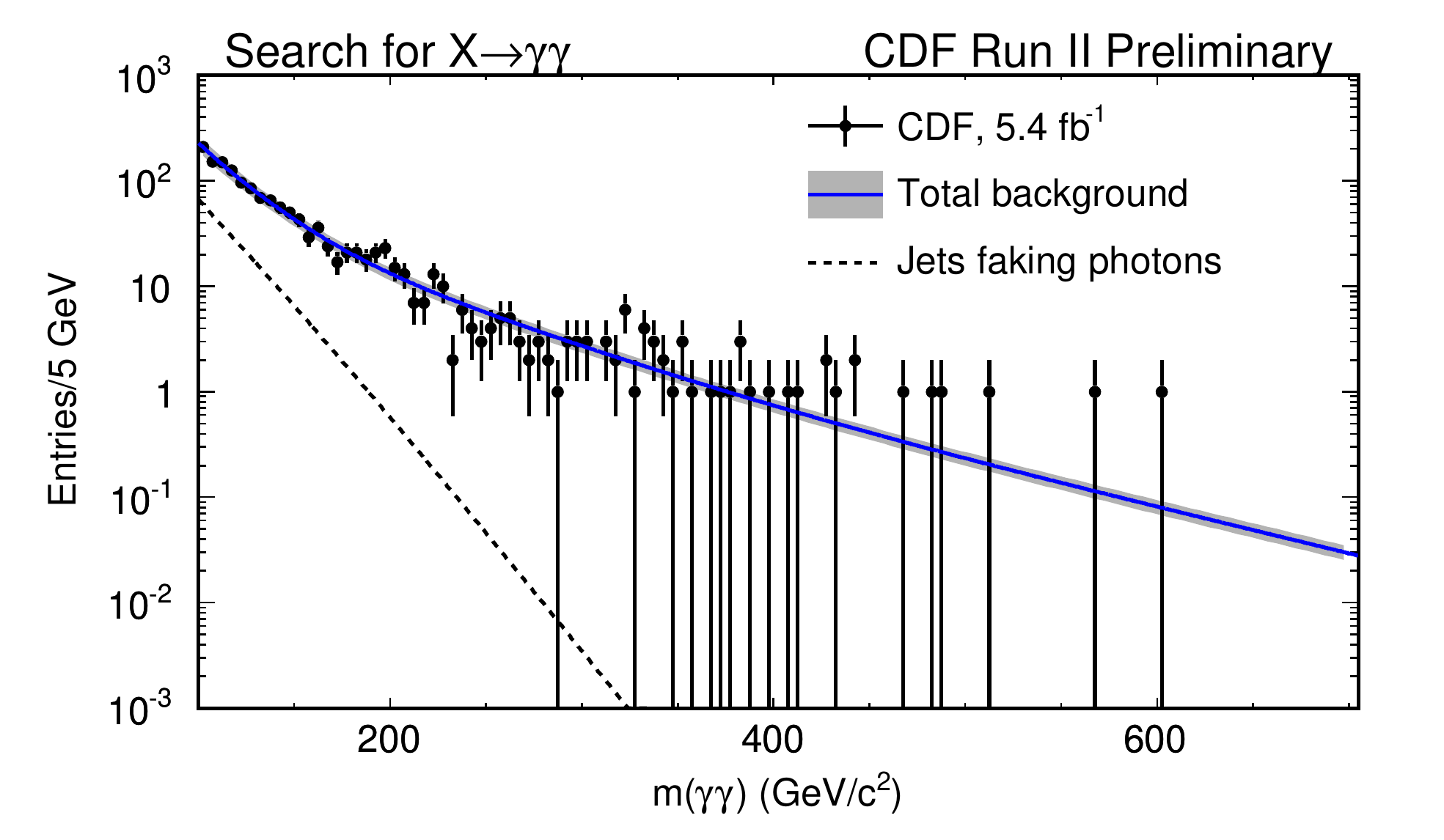}
\caption{Diphoton mass spectrum from the CDF search for RS-gravitons.}
\label{RSspectrum_gg}
\end{figure}

Figure \ref{RSspectrum_gg} shows the observed diphoton mass spectrum with the total background overlaid.
In the region around 200 GeV/$c^2$ there is an excess of events, with a local probability of such 
a fluctuation of 1.3\%.  This is consistent with the 1-sigma spread
of minimal probabilities in the range of 150-650 GeV/$c^2$, as determined with pseudoexperiments.

\section{Search for RS-gravitons decaying to dileptons}

CDF completed searches for the RS-gravitons in the dielectron \cite{RS_ee} and dimuon \cite{RS_mm}
channels using 5.7 fb$^{-1}$ of integrated luminosity.  Single-lepton triggers requiring an electron or muon
with transverse momentum $p_T>18$~GeV/$c$ are used, in addition to a 
trigger that remains efficient for $p_T>70$~GeV/$c$ electrons.  Offline two electrons (muons) are selected, at least one 
with energy ($p_T$) above 20 GeV (GeV$/c$).  The leptons come from the same primary vertex 
and are isolated, i.e., the excess energy in a cone $\Delta R =0.4$ around each lepton is less than 10\% of the 
energy (momentum) of the electron (muon).  Cosmic veto and photon-conversion removal are applied; no opposite charge is enforced.  

Main SM dilepton background to the RS-graviton dilepton signal is the Drell-Yan (DY) process 
$q\bar{q} \rightarrow Z/\gamma^* \rightarrow \ell^+ \ell^-$.  Significant background comes from 
QCD-originated events of one real lepton and one "fake" lepton, i.e., jet (track) faking an electron (muon).  
Minor backgrounds come from diboson ($WW$, $WZ$, $ZZ$, $W\gamma^*$) and $t\bar{t}$ processses with 
subsequent leptonic decays.  The QCD background is determined with CDF data, 
with the application of a probability that a jet (track) fakes an electron (muon) on events with one identified lepton.  
This probability is a function of the jet's (track's) transverse momentum and is determined by counting reconstructed
electrons and muons in a jet-rich CDF data, which is characterized by minimal leptonic content.  All other backgrounds are 
estimated with MC simulations ({\sc pythia}-generated events using {\sc cteq5l} PDF, processed by the CDF detector simulator) abosolutely 
normalized to the theoretical next-to-leading order cross sections, data luminosity, lepton-ID scale factors and trigger efficiencies.  

The main sources of systematic uncertainty on the MC-estimated backgrounds
are the theoretical cross sections (an 8\% effect on the event yields),
the luminosity (6\%), the lepton-ID efficiency (2\%), the parton distribution functions (2\%), and the trigger efficiency (0.5\%).
The total MC systematic uncertainty on the expected event yield is $\sim 10$\%.  The 
respective QCD-background systematic uncertainty is $\sim 50$\%, which comes from the variation
in the measurement of the fake probabilities using different jet-rich CDF datasets
triggered with varied jet-energy thresholds.  

\begin{figure}[!]
\begin{minipage}[!]{7.5cm}  
\begin{center}
\includegraphics[scale=.4]{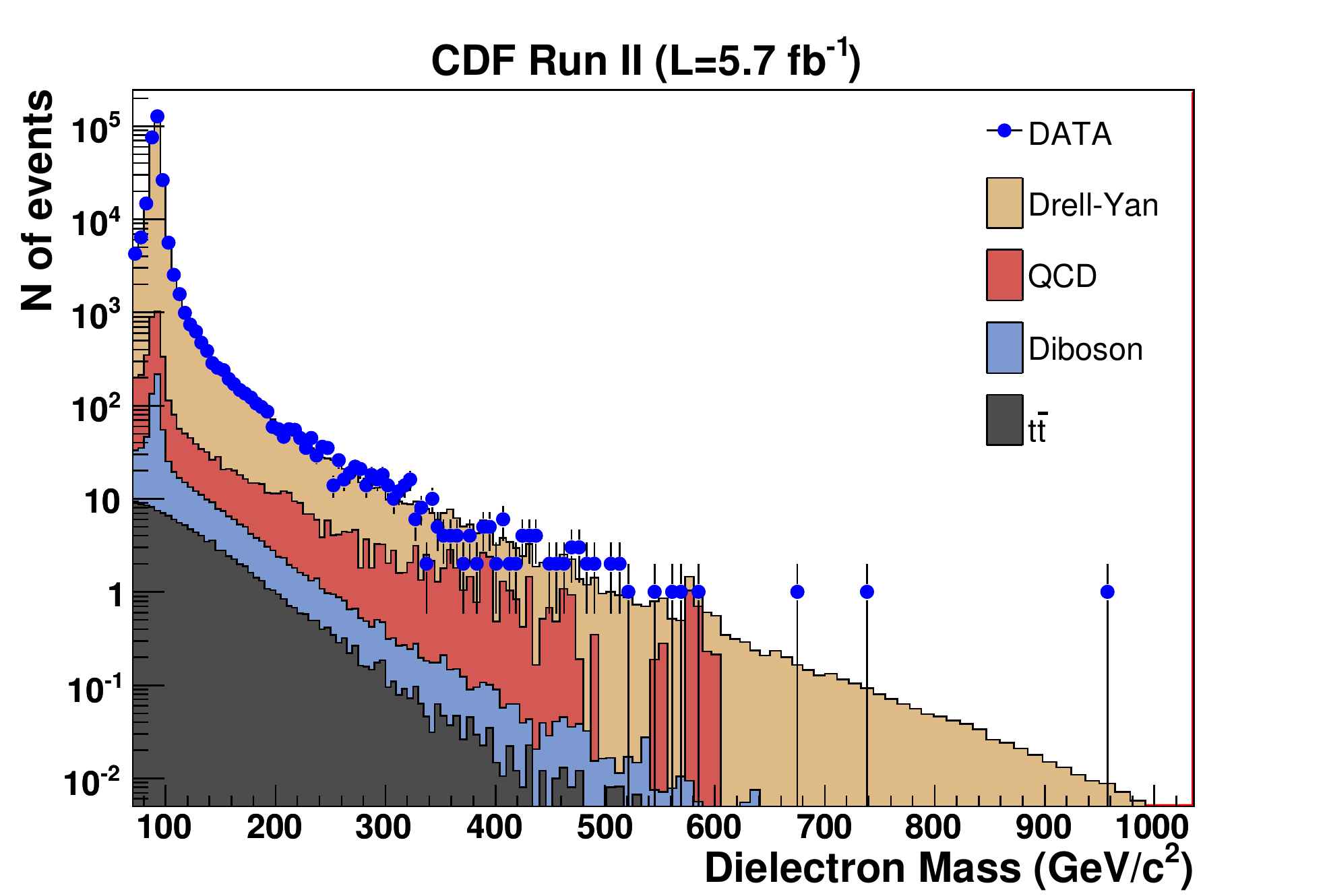}
\end{center}
\end{minipage}  
\hfill
\begin{minipage}[!]{7.5cm} 
\begin{center}
\includegraphics[scale=.4]{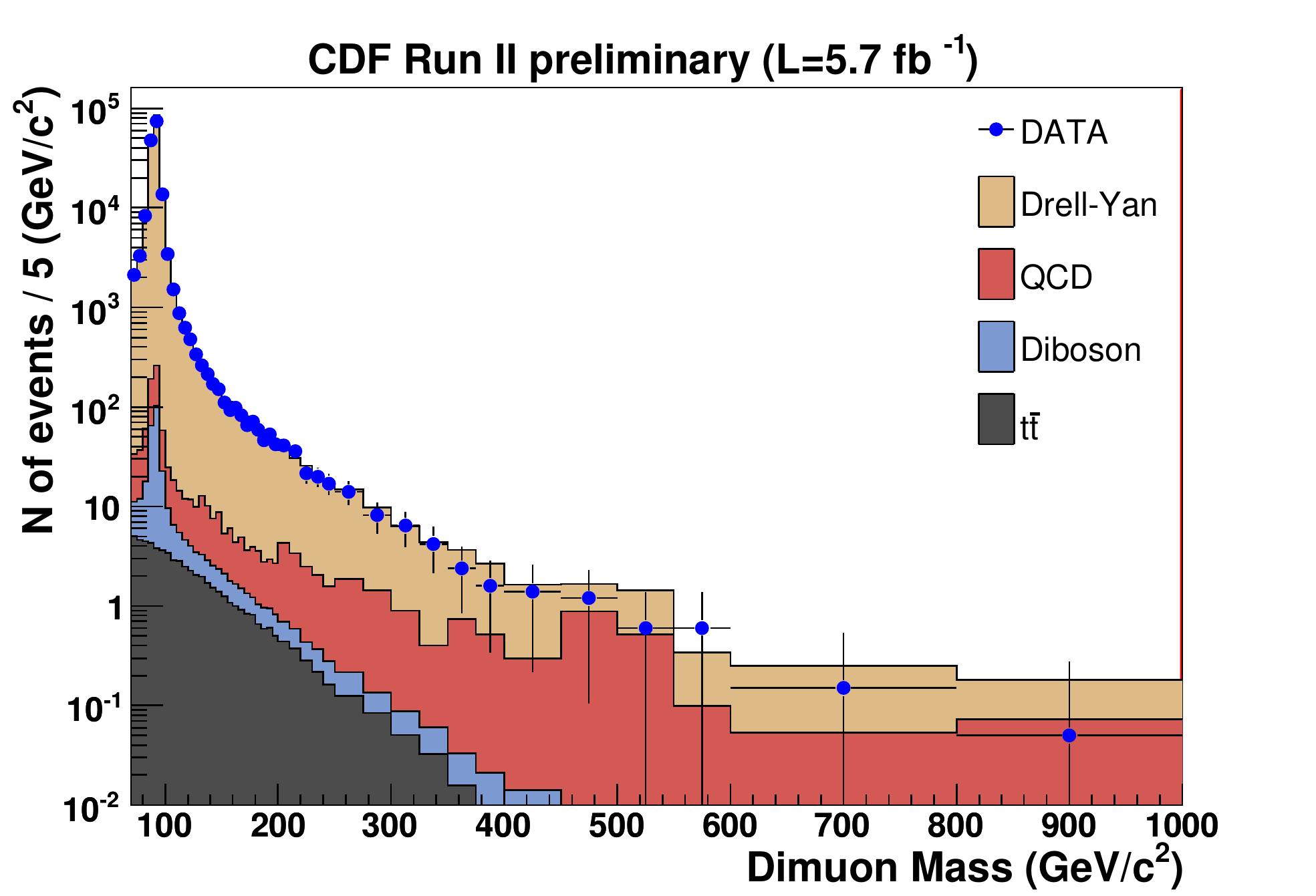}
\caption{Dielectron and dimuon mass spectra from the CDF search for RS-gravitons. \label{RSspectrum_ll}}
\end{center}
\end{minipage}  
\hfill
\end{figure}

The dielectron and dimuon mass spectra are consistent with expectation, as can be seen in Figure \ref{RSspectrum_ll}.  At the same time, the highest-dielectron-mass event ever observed at the time was detected ($M_{ee}=960$ GeV/$c^2$).   
In this event, the two electrons have transverse momenta of 482 and 468~GeV/$c$ and are oppositely charged.  The event is characterized
by very low hadronic activity (no jets are reconstructed) and by a low missing transverse energy 
of 17~GeV -- separated by 23 degrees in $\phi$ from the 468~GeV/$c$ electron -- coming most probably from the resolution
of the calorimeter.  The probability that at least one event is observed with a mass at least that high is 4\%.  This 
probability is within the 1-sigma range of minimal probabilities observed anywhere in the dilepton spectrum, as determined
by pseudoexperiments. 
\section{Combined Limits}
\begin{figure}[!]
\includegraphics[scale=.45]{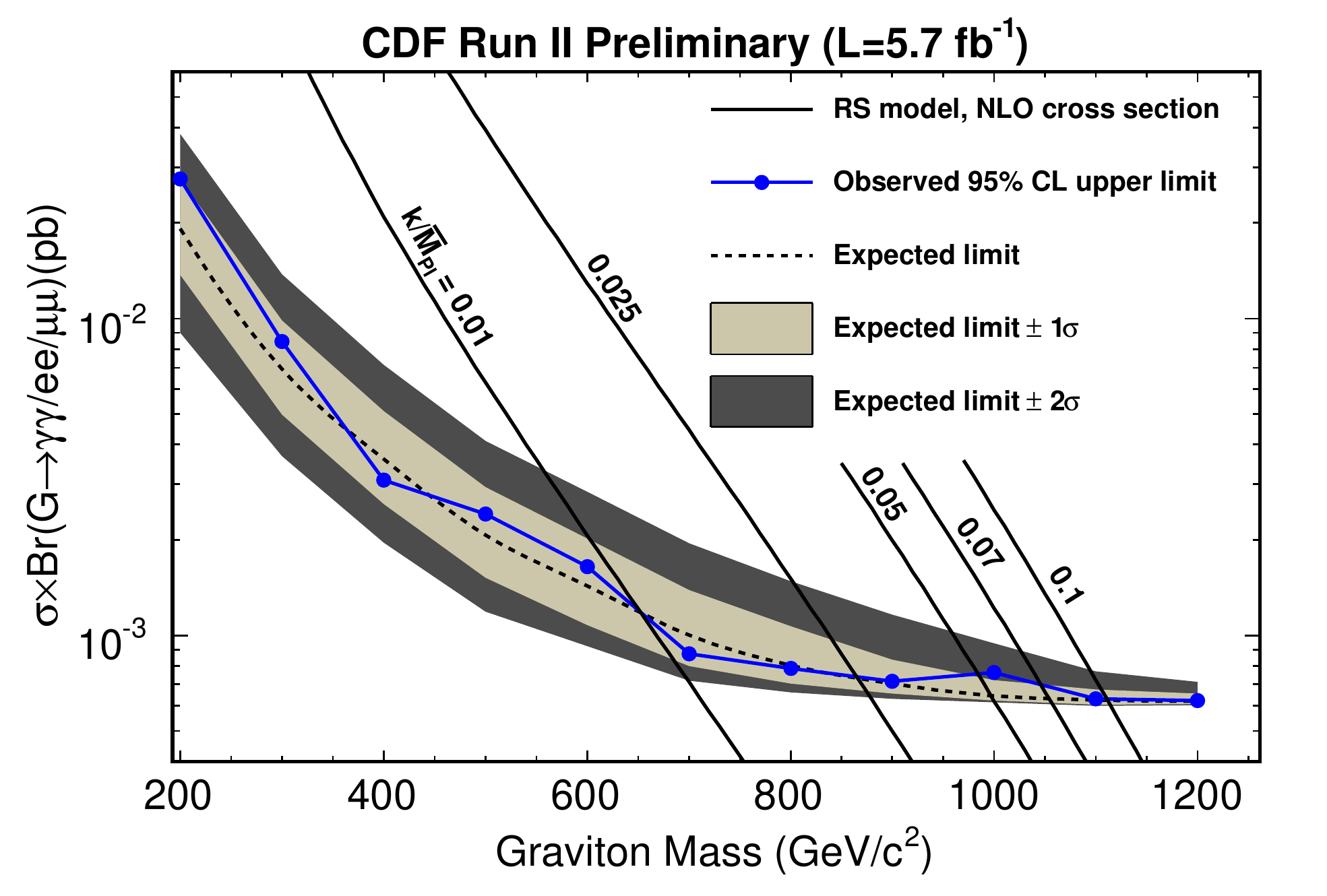}
\caption{The 95\% CL upper limit on the lightest RS-graviton production 
cross section, shown with theoretical cross sections for five values of $k/\overline{M}_{Pl}$. \label{RSlimit}}
\end{figure}
The measured diphoton and dilepton spectra can be used to set a limit on RS-graviton production \cite{RS_mm}.
Here we parametrize the RS model using the mass of the lightest RS graviton 
($m_1$) and the dimensionless parameter $\sqrt{8\pi} k/M_{Pl} \equiv k/\overline{M}_{Pl}$,
where $k$ is the curvature scale of the extra dimension and $M_{Pl}$ is the Planck mass. 
RS-graviton signal MC events are generated using {\sc pythia} and values of $m_1$ from 200 to 1100~GeV/$c^2$.   
The signal-MC events are generated and normalized in the same manner as the background-MC events.
The leading-order {\sc pythia} cross section is multiplied by a scale (``$K$-factor'') 
to correct for next-to-leading-order effects \cite{mathews}.

Figure \ref{RSlimit} shows the combined 95\% CL cross-section ($\sigma \times {\rm Br}(G\rightarrow \gamma\gamma/ee/\mu\mu)$) 
upper limit as a function of $m_1$ along with five theoretical cross-section 
curves for $k/\overline{M}_{Pl}=0.01$ to 0.1, a theoretically interesting range that would provide a solution to the hierarchy problem.  
The limits are set using a frequentist method that
compares the background-only with the signal+background hypotheses, taking into account correlated background systematic uncertainties across the channels \cite{tomjunk}.
The intersection of the cross-section exclusion limit 
with the theoretical cross-section curves gives the 95\% CL lower limit on $m_1$ for the respective coupling.
For $k/\overline{M}_{Pl}=0.1$, the $m_1$ lower limit is 1111~GeV/$c^2$, if we use proper mass-dependent 
RS-graviton $K$-factors, and 1141~GeV/$c^2$ assuming a fixed $K$-factor of 1.54, for comparisons with previous Tevatron results.
For $m_1>1$~TeV/$c^2$, cross sections greater than 0.6 fb are excluded at the 95\% CL.  At the time of release, 
these results were the most strigent in the world.




\end{document}